\documentclass[%
 reprint,
superscriptaddress,
 amsmath,amssymb,
 aps,prmaterials,
longbibliography,
floatfix
]{revtex4-1}
\usepackage{placeins}
\usepackage{graphicx}
\usepackage{dcolumn}
\usepackage{bm}
\usepackage{mathptmx}
\usepackage{booktabs} 
\usepackage{multirow}
\usepackage[table]{xcolor}

\begin{document}


\title{Phase Boundary Exchange Coupling in the Mixed Magnetic Phase Regime\\ of a Pd-doped FeRh Epilayer} 

\author{J.~R.~Massey}
\email{J.R.Massey@leeds.ac.uk}
\affiliation{School of Physics and Astronomy, University of Leeds, Leeds LS2 9JT, United Kingdom.}

\author{K.~Matsumoto}
\affiliation{Institute for Solid State Physics, University of Tokyo, Kashiwa, Chiba 277-8581, Japan.}

\author{M.~Strungaru}
\affiliation{Department of Physics, University of York, York YO10 5DD, United Kingdom.}

\author{R.~C.~Temple}
\affiliation{School of Physics and Astronomy, University of Leeds, Leeds LS2 9JT, United Kingdom.}

\author{T.~Higo}
\affiliation{Institute for Solid State Physics, University of Tokyo, Kashiwa, Chiba 277-8581, Japan.}

\author{K.~Kondou}
\affiliation{Center for Emergent Matter Science, RIKEN, Wako, Saitama 351-0198, Japan.}

\author{R.~F.~L.~Evans}
\affiliation{Department of Physics, University of York, York YO10 5DD, United Kingdom.}

\author{G.~Burnell}
\affiliation{School of Physics and Astronomy, University of Leeds, Leeds LS2 9JT, United Kingdom.}

\author{R.~W.~Chantrell}
\affiliation{Department of Physics, University of York, York YO10 5DD, United Kingdom.}

\author{Y.~Otani}
\affiliation{Institute for Solid State Physics, University of Tokyo, Kashiwa, Chiba 277-8581, Japan.}
\affiliation{Center for Emergent Matter Science, RIKEN, Wako, Saitama 351-0198, Japan.}

\author{C.~H.~Marrows}
\email{c.h.marrows@leeds.ac.uk}
\affiliation{School of Physics and Astronomy, University of Leeds, Leeds LS2 9JT, United Kingdom.}

\date{\today}

\begin{abstract}
Spin-wave resonance measurements were performed in the mixed magnetic phase regime of a Pd-doped FeRh epilayer that appears as the first-order ferromagnetic-antiferromagnetic phase transition takes place. It is seen that the measured value of the exchange stiffness is suppressed throughout the measurement range when compared to the expected value of the fully ferromagnetic regime, extracted via the independent means of a measurement of the Curie point, for only slight changes in the ferromagnetic volume fraction. This behavior is attributed to the influence of the antiferromagnetic phase: inspired by previous experiments that show ferromagnetism to be most persistent at the surfaces and interfaces of FeRh thin films, we modelled the antiferromagnetic phase as  forming a thin layer in the middle of the epilayer through which the two ferromagnetic layers are coupled up to a certain critical thickness. The development of this exchange stiffness is then consistent with that expected from the development of an exchange coupling across the magnetic phase boundary, as a consequence of a thickness dependent phase transition taking place in the antiferromagnetic regions and is supported by complimentary computer simulations of atomistic spin-dynamics. The development of the Gilbert damping parameter extracted from the ferromagnetic resonance investigations is consistent with this picture. \end{abstract}

\maketitle

\section{Introduction}

B2-ordered FeRh undergoes a first order metamagnetic phase transition from an antiferromagnet (AF) to a ferromagnet (FM) on heating through a transition temperature $T_\text{T} \sim 380$~K \cite{Fallot1939}. The proximity of the transition to room temperature and its sensitivity to external stimuli \cite{Staunton2014,Kouvel1966,Barua2013,Wayne1968,Annaorazov1996,Maat2005} make it an excellent candidate for use in possible magnetic memory devices architectures \cite{Thiele2003,Cherifi2014,Lee2015}, including those involving AF spintronics \cite{Marti2014,Moriyama2015}. Coexistence of the two magnetic phases whilst the material undergoes the transition has been well studied using various magnetic imaging techniques \cite{Baldasseroni2012,Kinane2014,devries2014,Baldasseroni2015,Almeida2017,Temple2018}. This region of the transition is known as the mixed magnetic phase (MMP) and has AF material in direct contact with FM material, meaning there is potential for an exchange coupling between the two phases. Any interphase exchange coupling may affect the transition kinetics and the performance of FeRh based devices \cite{Moriyama2015}.

The properties of this exchange coupling remain elusive. There have been claims that it has been measured using ferromagnetic resonance (FMR) experiments, which saw a change in the anisotropy field that develops when entering the MMP from the fully FM regime \cite{mancini}. However, this behavior has also been attributed to the field due to magnetoelastic effects at the film/substrate interface \cite{Kumar2018} and general phase coexistence \cite{Heidarian2017}. More recently, vertical exchange bias and enhanced coercivity, indicative of an interphase exchange coupling, has been seen in FeRh \cite{Gray2019}. The direct study of the influence of any such coupling in the GHz regime is of importance for the design of FeRh based spintronic devices that are expected to operate on ns timescales.

Spin-wave resonance (SWR) is an extension of FMR in which higher-order non-uniform modes are studied. The fact that the magnetization flexes out of a non-uniform state allows for the extraction of the exchange stiffness $A$ of the magnetic material including magnetic multilayers \cite{Goennenwein2003,Klinger2015,Yin2017,coeyswr,vanStapele1985}. In SWR, pinning conditions allow for the excitation of perpendicular standing spin-waves (PSSWs) for external magnetic fields applied perpendicularly to the film surface \cite{coeyswr}. The frequency of the PSSW of mode number $n$, $f_n$, is determined by the exchange stiffness across the film thickness $t$ such that \cite{vanStapele1985,Sipr2019},
\begin{equation} \label{swr1}
f_n = f_\text{FMR} + \frac{2 A g \mu_\text{B}}{M_\text{S}h} \left( \frac{n \pi}{t} \right) ^2,
\end{equation}
where $f_\text{FMR}$ is the $n = 0$ FMR mode frequency, $g$ is the spectroscopic splitting factor extracted from the behavior of the FMR mode, $h$ is the Planck constant and $M_\text{S}$ is the saturation magnetization \cite{coeyswr}.

Here, we present SWR investigations on a Pd-doped FeRh epilayer within the MMP regime. The Pd doping was used to reduce $T_\text{T}$ to an experimentally convenient value. We found that the measured value of $A$ is suppressed compared to the expected value of the FM phase for even the slightest deviation in the phase volume fraction from a fully FM state. This is attributed to the influence of the AF phase, which is seen to contribute to the measured exchange stiffness. The behavior of the exchange stiffness within the AF layer is shown, using computer simulations of atomistic spin dynamics, to correspond to the development of the exchange coupling across the magnetic phase boundary as a consequence of a thickness dependent phase transition (TDPT) in the AF layer. These findings are also supported by the behavior of the Gilbert damping parameter extracted from the FMR measurements.

\section{Sample Growth and Characterization}

\begin{figure*}[t!]
  \includegraphics[width = 16cm]{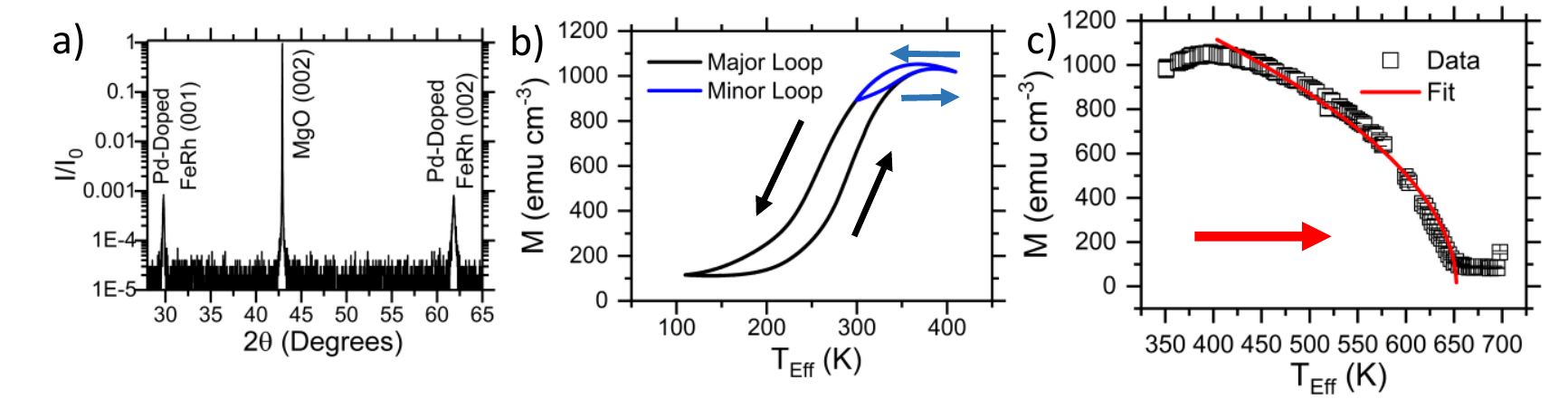}
  \caption{Pd-doped FeRh film characterization. (a) XRD spectrum with indexed Bragg peaks. (b) Magnetization of the Pd-Doped FeRh sample between 100 and 400~K on the major loop (black line) and between 290 and 400~K on the minor loop (blue line), both measured with a 1~T field applied in the film plane. (c) Higher temperature magnetization measurement (circles) alongside the fitting used to extract $T_\mathrm{C}$ (line) measured in a 0.1~T field applied in the film plane. The arrows depict the temperature sweep direction. The effective temperature $T_\text{Eff}$ accounts for the applied field as described in the text. \label{fig:XRD}}
\end{figure*}

The B2-ordered Pd-doped FeRh epilayer was grown using DC magnetron sputtering. The MgO substrate was annealed overnight at 700$^{\circ}$C, the sample was then deposited at a substrate temperature of 600$^{\circ}$C and annealed \textit{in situ} at 700$^{\circ}$C for 1 hour. 3\% Pd doping on the Rh site was used to lower $T_\text{T}$ so the transition spanned room temperature \cite{Kouvel1966,Barua2013} to match the capabilities of the measurement apparatus. The sample thickness was measured using x-ray reflectivity as $t = 134 \pm 4$~nm. X-Ray diffraction (XRD), shown in Fig.~\ref{fig:XRD}(a), shows the $(001)$ and $(002)$ Pd-doped FeRh reflections astride the central substrate peak, which confirms the presence of an epitaxial growth having a B2 order parameter $S = 0.76 \pm 0.02$ \cite{warren,leGraet2013,devries2013} and a lattice constant $a = 2.998 \pm 0.001$~\AA.

The magnetic behavior of the sample was measured using a SQUID-vibrating sample magnetometer (VSM) between 100 and 400~K in a 1~T field that was applied in the film plane, and is shown in Fig.~\ref{fig:XRD}(b). The black curve shows the major loop where the transition is completed in both directions. As it was not possible to cool below room temperature during the SWR experiment, a minor loop taken through the available temperature range (290-400~K) is also shown by the blue line, for comparison with the SWR data. The temperature scale used here is the effective temperature, $T_\text{Eff}$, which is used as the measure of position within the transition as external magnetic fields affect $T_\text{T}$ \cite{Kouvel1966,Maat2005,Annaorazov1996,Barua2013}. $T_\text{Eff}$ is calculated via $T_{\text{Eff}} = T_0 - \frac{d T_\text{T}}{d(\mu_0 H_\text{Ext})} \mu_0 H_\text{Ext}$, where $T_0$ is the measured sample temperature and $dT_\text{T}/ d(\mu_0 H_\text{Ext}) = -(9.6 \pm 0.6)$~KT$^{-1}$ when cooling and $dT_\text{T}/ d(\mu_0 H_\text{Ext}) = -(9.3 \pm 0.5)$~KT$^{-1}$ when heating. These values were measured by tracking the transition midpoint of our film through a range of fields \cite{Maat2005}. The saturation magnetization at the temperature when the transition is completed is $\mu_0 M_\text{S} = 1.32 \pm 0.05$~T, with a moment per Fe atom of $\mu_\text{Fe} = 3.1 \pm 0.1  ~ \mu_\text{B}$.

Fig.~\ref{fig:XRD}(c) shows the magnetization at higher temperatures which is fitted to $M = M_0 \left[ 1- \left( T/T_\text{C} \right) \right]^\beta$ to extract the Curie temperature, $T_\text{C} = 652 \pm 1$~K with $\beta = 0.51 \pm 0.02$, which gives good agreement with the mean-field model, which predicts $\beta = 0.5$ \cite{mftbeta}. We then calculated the exchange constant for the fully FM regime, $J_\text{FM}$, using the mean field model via $J_\text{FM} = 3 k_\text{B} g^2 T_\text{C}/2\mu_\text{Fe}^2 Z  = (1.01 \pm 0.05)\times 10^{-21}$ J \cite{Stohrbook3}. For the purposes of this calculation it was not possible to measure the value of $g$ for the fully FM phase and so we took $g = 2.05 \pm 0.06$, as measured previously \cite{mancini}. It has been shown previously using computer simulations of atomistic spin-dynamics that the metamagnetic transition in B2-ordered FeRh can be replicated by modelling only the Fe moments and treating their interactions due to the Rh moment as higher-order exchange interactions between quartets of adjacent Fe atoms \cite{Barker2015}. FeRh therefore can be modelled as a simple cubic structure comprising only Fe atoms, which makes $Z = 6$. It is then possible to convert this to an exchange stiffness for the fully FM regime, $A_\text{FM}$, using the following relation $A_\text{FM} = J_\text{FM}S^2 / a = 7.5 \pm 0.6$~pJm$^{-1}$, where $S$ is the spin per atom calculated from $\mu_0 M_\text{S}$ \cite{coeyA}.

\section{Spin-Wave Resonance Measurements}

For the SWR measurements the sample was placed face down on a two-port coplanar waveguide with a ceramic heater used for temperature control. The external magnetic field, $\mu_0 H_\text{Ext}$ was applied perpendicular to the film surface and transmission through the waveguide was measured using a vector network analyzer. The transmission through the waveguide is measured using either $S_\text{21}$ or $S_\text{12}$ and is presented as $I = S_{ij}(\mu_0 H_\text{Ext}) - S_{ij}(\mu_0 H_\text{Ext} = 0~\text{T})$, where $S_{ij}$ is the magnitude of the trace. As the transition temperature in FeRh is sensitive to the application of $\mu_0 H_\text{Ext}$ \cite{Kouvel1966,Maat2005,Annaorazov1996,Barua2013}, the frequency was swept between 0.01 to 26~GHz to identify the resonance positions, whilst $\mu_0 H_\text{Ext}$ was held constant at 50~mT intervals between 1.4 and 2~T. These measurements were performed at various temperatures on both the heating and cooling branches of the transition from the close to the fully FM state ($T_\text{Eff} \sim 360$~K) down to where SWR modes could no longer be observed ($T_\text{Eff} \sim 310$~K). The available temperature range in this experiment is between 301 and 338~K and was limited by the experimental apparatus.

\begin{figure}[t!]
  \includegraphics[width = 7cm]{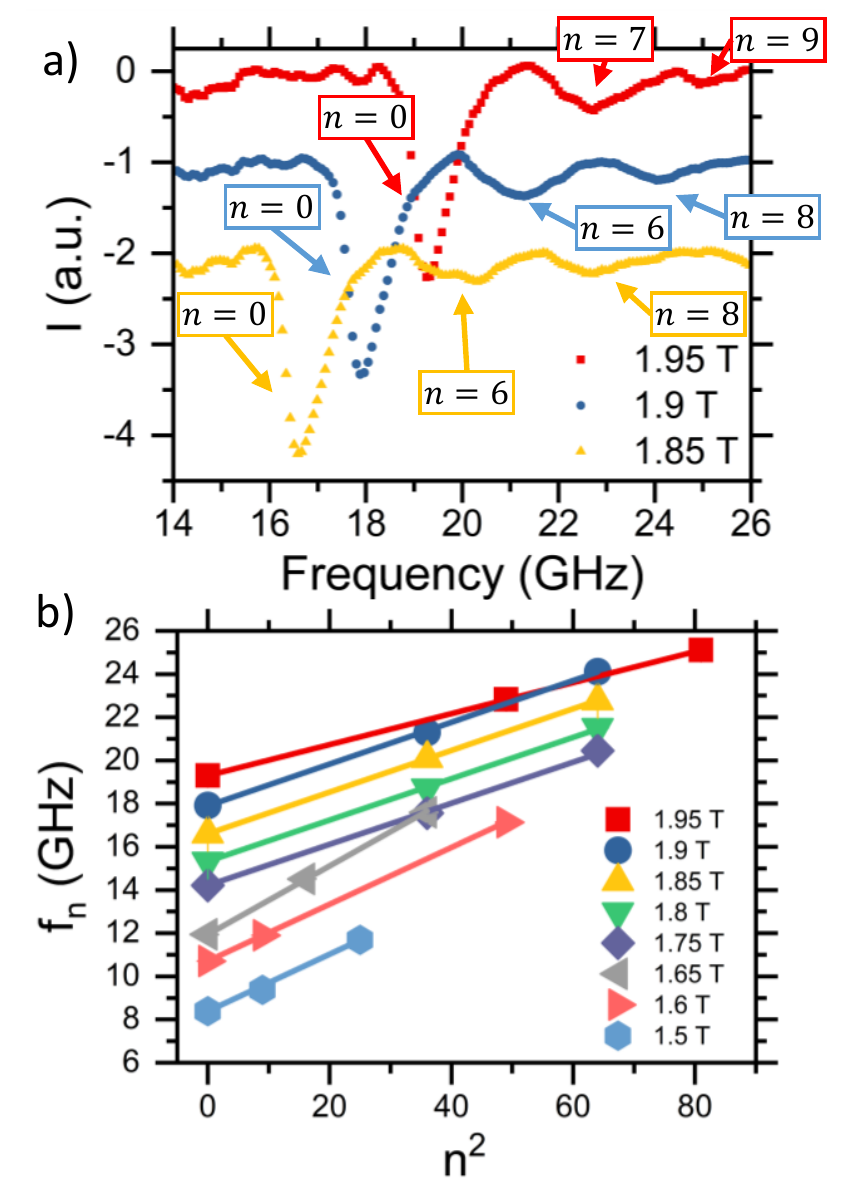}
  \caption{SWR measurements. (a) example SWR spectra acquired at $T_0 = 338.4$~K on the heating branch, which show the presence of extra modes attributed to the excitation of PSSW modes, which are labelled accordingly.  (b) frequency of the $n$th mode excitation, $f_\text{n}$, plotted against the square of the assigned mode number, $n^2$. The solid lines in this figure are linear fits used to extract $A$ in accordance with Eq.~\ref{swr1}.\label{fig:SWR}}
\end{figure}

Fig.~\ref{fig:SWR}(a) shows an example set of SWR frequency spectra taken at $T_0 = 338.4$~K on the heating branch, which are labeled with the assigned PSSW mode number. They show prominent $n = 0$ FMR modes along with higher frequency modes corresponding to SWR excitations. Fig.~\ref{fig:SWR}(b) shows the frequency for the PSSW excitation for a given mode number, $f_\text{n}$, against $n^2$ for the spectra in panel (a) of the same figure. The solid lines in this figure are linear fits to the data that are used, along with Eq.~\ref{swr1}, to extract $A$ for each value of $\mu_0 H_\text{Ext}$ in a measurement set, which is then converted to values of $T_\text{Eff}$. The linear relationship between $f_n$ and $n^2$ seen in Fig.~\ref{fig:SWR}(b) confirms the validity of equation \ref{swr1} to explain the behavior here.

In SWR, the mode number of the excited PSSWs is very dependent on the boundary conditions acting on the spin wave modes within the system \cite{Seavey1958,Schmool1998,Schoen2015}. In our FeRh epilayer, there would be boundary conditions to satisfy at either end of the film thickness, as well as those present at any interface between the two magnetic phases. This makes the exact boundary conditions difficult to discern. To account for this we assumed that PSSWs with both odd and even mode numbers are capable of being excited in this system. These conditions have been previously applied to systems in which a uniform magnetization with asymmetric pinning conditions at either end of the film thickness \cite{Schoen2015}, as well as systems that are exchange coupled at the interface between two layers \cite{Qin2018}. Equation~\ref{swr1} states that when the mode numbers are correctly assigned that $(f_n - f_\text{FMR})/n^2$ is constant. This condition, alongside checking the distance between adjacent PSSW modes against the expected distance for a set of mode numbers was used to assign the mode numbers. In Fig.~\ref{fig:SWR}(b) it is clear that not all the assigned mode numbers are consecutive and there are some that are missing. This is attributed to the fact that the model used in this work may not accurately reflect the exact pinning conditions of the FeRh system.

\begin{figure}[t!]
  \includegraphics[width = 7cm]{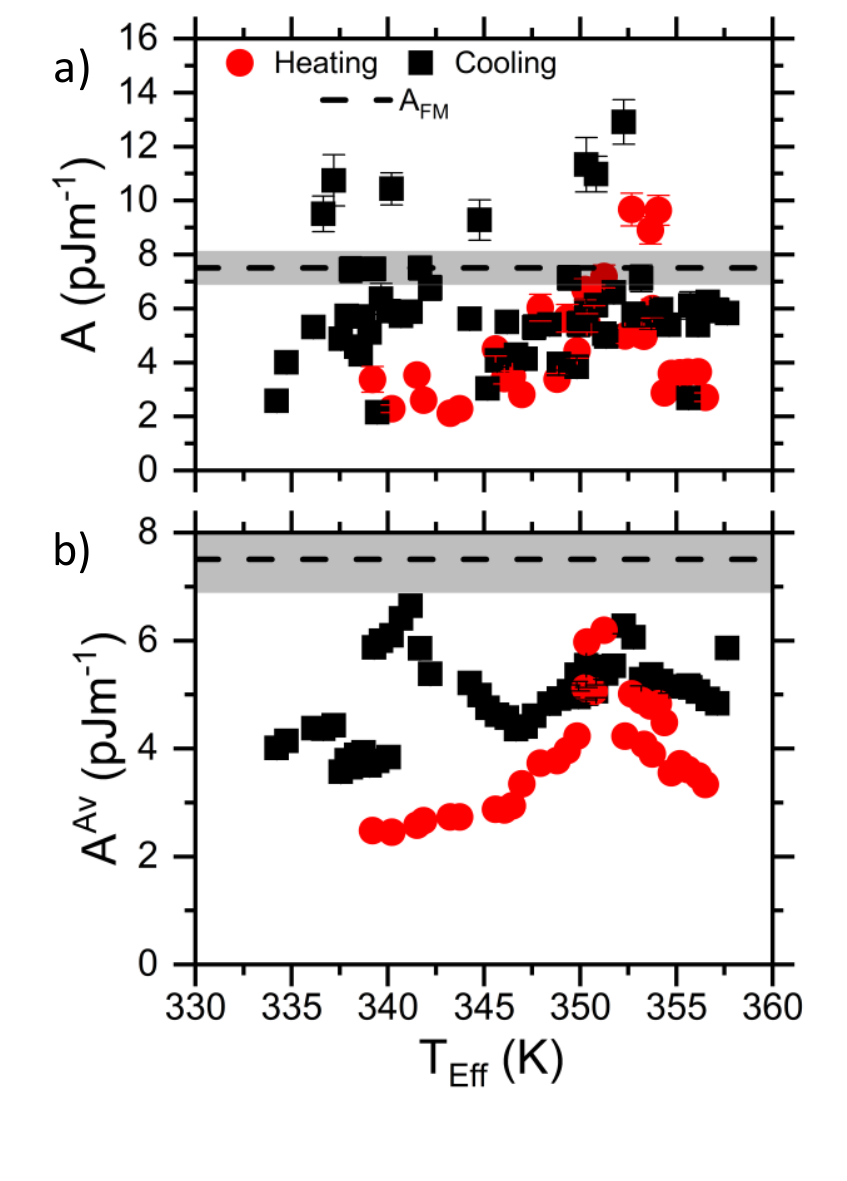}
  \caption{Exchange stiffness. (a) extracted values of $A$ obtained for all measurements plotted against $T_\text{Eff}$. Measurements performed on the cooling branch of the transition are shown by the black squares, whilst the heating branch measurements are shown using red circles. This convention is adopted throughout this work and should be assumed to be the case unless specified otherwise. The black dashed line in this figure shows the position of $A_\text{FM}$, with the region encompassed by the error bars shown in grey. (b) nine-point symmetric rolling adjacent weighted average of $A$, $A^\text{Av}$, also plotted against $T_\text{Eff}$. \label{fig:A}}
\end{figure}

The dependence of $A$ on effective temperature $T_\text{Eff}$ is shown in Fig.~\ref{fig:A}(a). In this calculation of $A$ it is assumed, for reasons explained later, that the PSSW excitations span the entire film thickness. The values of $A$ are calculated using equation \ref{swr1} and the data shown in Fig.~\ref{fig:SWR}(b) for all measurements where the PSSW peaks are discernible from the background. Due to the large spread in the data for $A$, a symmetric rolling nine-point weighted average was performed to smooth the data. The behavior of the smoothed exchange stiffness, $A^\text{Av}$, is shown against $T_\text{Eff}$ in Fig.~\ref{fig:A}(b). From this figure it can be seen that there are two peaks in $A^\text{Av}$ when cooling at $T_\text{Eff} \sim 340$ and 350~K and one when heating at $T_\text{Eff} \sim 350$~K. These peaks are superimposed upon an overall decrease in $A^\text{Av}$ with decreasing $T_\text{Eff}$ across the measurement range.

It is also clear here that the values of $A^\text{Av}$ consistently fall short of the value of $A_\text{FM}$. It is important to note that the SWR apparatus was tested by performing measurements on a permalloy (Ni$_{0.81}$Fe$_{0.19}$) film, yielding a value of $A = 10.6 \pm 0.2$ pJm$^{-1}$, which is consistent with the value obtained by Schoen \textit{et al}. \cite{Schoen2015}, when using the value of $g$ extracted from Shaw \textit{et al}. \cite{Shaw2013}. Again, in the calculation of $A$ for Py, the same method was used to assign the mode numbers as that used for the FeRh system, and mode numbers are seen to be missing there also, which adds credence to the hypothesis that this behavior is real. All of the measurements presented in Fig.~\ref{fig:A} take place in regions of the transition away from the fully FM state, and so some volume fraction, however small (see Fig.~\ref{fig:XRD}(b)), is in the AF phase in every measurement. Therefore, this deviation from $A_\text{FM}$ for all measurements implies that the AF phase has a profound effect on the measured value of $A$ for this material system. In order to identify why this is the case, a model of how the two magnetic phases develop relative to each other through the transition is required.

\begin{figure}[t!]
  \includegraphics[width = 7cm]{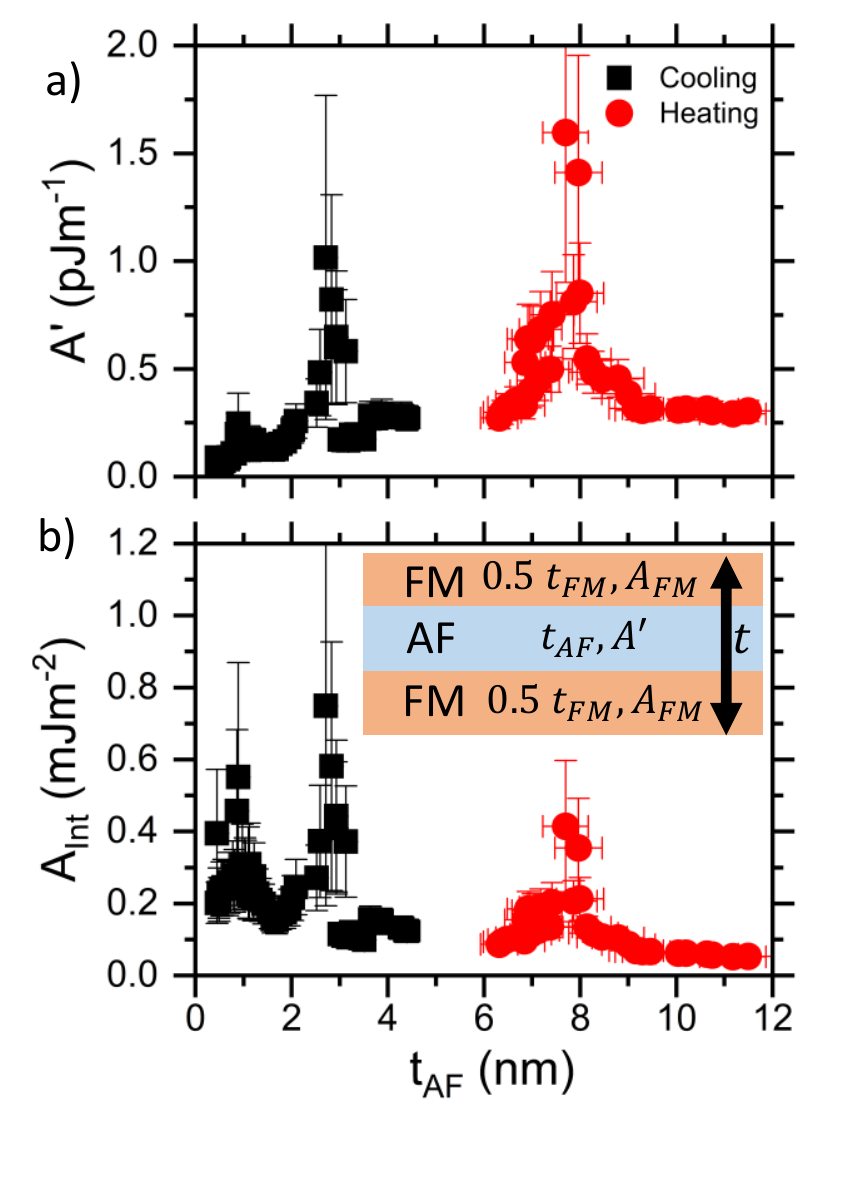}
  \caption{Application of the trilayer model to the SWR. (a) the development of the exchange stiffness in the AF region, $A^\prime$, against the AF layer thickness calculated using the trilayer model, $t_\text{AF}$. (b) the development of the interfacial exchange coupling, $A_\text{Int}$, plotted against $t_\text{AF}$. Inset into panel (b) is a schematic of the trilayer model where the FM layer is split into two layers of equal thickness. Within this schematic, the properties of each region are described using their thickness, $t_\text{i}$, and their exchange stiffness, $A_\text{i}$.\label{fig:trilayer}}
\end{figure}

To do this we have developed a simple trilayer model. Photoemission electron microscopy (PEEM) has revealed that the top surface of FeRh can be fully ferromagnetic at a temperature where the sample as a whole is not in the fully FM regime \cite{Baldasseroni2012,Baldasseroni2015}. This therefore means that the transition has completed on the top surface, but not in the bulk of the material. Such behavior has also been seen using electron holography, which demonstrated that either surface has a lower transition temperature than the bulk of the material, and so the FM phase nucleates there and permeates into the bulk of the material as the transition progresses \cite{Gatel2017}. This picture is also consistent with grazing incidence x-ray diffraction studies of the structural component of the transition \cite{Kim2009} and polarized neutron reflectometry that shows FM order is retained at the film interfaces whilst the interior is AF \cite{Fan2010}. In the saturating out-of-plane fields applied in the experiment, the non-uniform in-plane domain structure observed by PEEM will not exist. Therefore, we model the MMP as two layers of FM material that are separated by a layer of AF material and the system can be described as a FM/AF/FM trilayer. A schematic of the model is shown in the inset of Fig.~\ref{fig:trilayer}(b). In the thermal hysteresis regime the phase volume of the FM phase can be calculated using $\phi(T_\text{Eff}) = M(T_\text{Eff})/M_\text{S}$ based in the data in Fig.\ref{fig:XRD}(b). Using this model, each of the layers can be described by their thickness. FM domains extend from either surface with total thickness $t_\text{FM}= \phi t$, whilst the AF region has a thickness $t_\text{AF} = (1-\phi)t$.

The possibility now arises that the behavior we see here could be due to the confinement of PSSW excitations within these thinner FM layers, changing the value of $t$ in Eq.~\ref{swr1}. As $t_\text{FM} \propto \phi$ and $\phi$ decreases across the temperature range here when cooling, according to equation \ref{swr1} the measured value of $A$ would increase across the temperature range. Conversely, when heating $\phi$ would increase across the temperature range here meaning the measured value of $A$ would decrease. This is clearly not consistent with our data. Thus we expect that PSSW excitations still span the entire film thickness $t$, and so must travel through the AF layer, which then contributes to the measured behavior. Thus the use of the whole thickness $t$ in the initial calculation of $A$ is justified. This also provides further justification for the trilayer model in that it can be seen to be consistent with another important feature of the data here, the fact that the SWR signal is lost when there is still a substantial fraction $\phi \approx 0.9$ of the material in the FM phase. The portion in the AF phase must laterally span the sample in order for there not to be any regions where a PSSW can vertically span the entire film. A thin AF layer is the only way to do this.

SWR in magnetic multilayers has been considered before by van Stapele \textit{et al}. \cite{vanStapele1985}. In that work it is stated that in magnetic multilayers with interfacial exchange coupling, when the system is driven at the resonant frequency of one of the layers but not the other, the measured value of $A$ becomes an effective measure, $A_\text{Eff}$, across the whole thickness \cite{vanStapele1985}. In the van Stapele model, the dormant layers mediate an exchange coupling between the layers driven at resonance, which means $A_\text{Eff}$ now consists of two contributions: the exchange stiffness of the layer driven at resonance and the exchange stiffness of the dormant layer and $A_\text{Eff}$ becomes a volume weighted average of the exchange stiffness for each of the different layers \cite{vanStapele1985}. In this scenario, all that is required of the dormant layer is that it possesses an exchange stiffness in order to carry the excitation through the stack \cite{vanStapele1985}. This means that the same model can be applied to the case here, where the layer not driven at its resonant frequency is AF.

By taking our measured value $A^\text{Av} = A_\text{Eff}$ and decomposing this average it is possible to calculate the exchange stiffness of the AF layer, $A^\prime$. It is important to note that this is not a ferromagnetic exchange stiffness \textit{per se}, but rather a measure of the rigidity in the spin structure that allows spin waves to propagate. To do this calculation, the FM regions were assigned the value of $A$ measured for the fully FM phase, $A_\text{FM}$, and both magnetic phases have their thicknesses determined by the trilayer model. In the small wavevector limit, it is possible to calculate $A^\prime$ using \cite{vanStapele1985},
\begin{equation}\label{eq:SWR_Jint}
A^\prime = \frac{A_\text{Eff}A_\text{FM}(1-\phi)}{A_\text{FM} - \phi A_\text{Eff}},
\end{equation}
the results of which are plotted against $t_\text{AF}$ in Fig.~\ref{fig:trilayer}(b). Interestingly, when cooling, the value of $A^\prime$ rises as $t_\text{AF}$ increases from 0 to 1~nm. This increase then plateaus up to $t_\text{AF} \sim 2$~nm, after which $A^\prime$ peaks at $t_\text{AF} \sim 3$~nm. A peak is also seen on the heating branch of the transition at $t_\text{AF} \sim 8$ nm. This is also seen to be sitting on a constant background, which is higher than that seen on the cooling branch. The non-zero value of $A^\prime$ indicates that the AF layer has an exchange stiffness that develops with $t_\text{AF}$ and contributes to the behavior seen here.

In the van Stapele model, the exchange stiffness of the layer not driven at its resonant frequency, the AF layer in this case, is responsible for the determination of the interfacial exchange coupling strength $A_\text{Int}$, such that $A_\text{Int} = 2A^\prime/t_\text{AF}$ \cite{vanStapele1985}. The behavior of $A_\text{Int}$ through the measurement range is shown plotted against $t_\text{AF}$ in Fig.~\ref{fig:trilayer}(b). $A_\text{Int}$ is non-zero through the measurement range here and decays in a manner consistent with that expected for an interfacial exchange coupling. Interestingly, superimposed on this characteristic decaying background, there are peaks in the extracted value of $A_\text{Int}$ that align with those seen in $A^\prime$.

The thickness dependence of $A^\prime$ has been seen before in the context of exchange bias, since it is reminiscent of the development of an exchange field exerted by an AF layer on a FM layer. There it is well understood in terms of the development of AF order in ultrathin films with increasing thickness \cite{Ali2003,Khan2018,Mccord2004}. At the lowest thicknesses, up to 1 or 2~nm, there is no effect of the AF layer on the magnetic properties of the FM, since the material is too thin to possess AF order and presents instead in a paramagnetic (PM) state. However, as the thickness increases the phenomena associated with exchange bias set in, indicating that stable AF-ordered material has appeared, with a characteristic peak in the exchange bias field, often explained in terms of a random field model \cite{Malozemoff1987,Malozemoff1988}, which is similar to that seen here in the AF thickness dependence of $A^\prime$. In the case of exchange bias, this behavior can be described by a TDPT between a PM and AF state that takes place in the AF layer, raising the question of whether a similar mechanism might be at play here. The development of an exchange coupling predicted by such a phase transition is consistent with the behavior of $A^\prime$ seen in this experiment between 0 and 2 nm when cooling. In this case however, FeRh does not exhibit a PM phase between AF and FM when undergoing its phase transition \cite{Stamm2008}. Nevertheless, our results suggest that when FM order starts to be lost, the AF layer is unable to establish full AF order until $t_\text{AF}$ exceeds a critical thickness that is at the position of the peak in $A^\prime$. The origin of the peak in $A^\prime$ seen at around $t_\text{AF} \sim 3$ nm is unclear at this stage.

\begin{figure}[t]
  \includegraphics[width = 7cm]{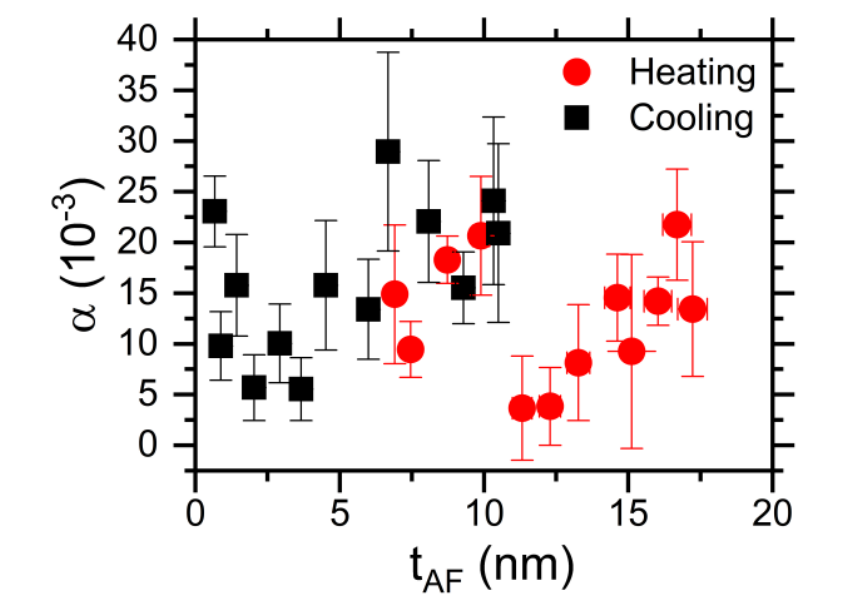}
  \caption{Gilbert damping parameter measured from FMR for both transition branches.}
\label{fig:damping}
\end{figure}

Evidence for phase transitions taking place in AF materials has been seen from the presence of peaks in the Gilbert damping parameter, $\alpha$, when passing through the transition \cite{Mccord2004,Frangou2016}. This has been seen for the TDPT in the Py/IrMn system \cite{Mccord2004}, as well as the transition from an AF to a PM state at the N\'eel temperature of IrMn in the same material system \cite{Frangou2016}. To see if the same behavior is seen here, the behavior of $\alpha$ extracted from the experimental FMR mode is shown against $t_\text{AF}$ in Fig.~\ref{fig:damping}. After fitting a Lorentzian profile to the FMR peaks to extract the linewidth, $\Delta f$ and the FMR frequency, $f_\text{FMR}$, it is possible to extract $\alpha$ using the following relation,
\begin{equation}
  \Delta f = \Delta f_0 + 2\alpha f_\text{FMR},
\end{equation}
where $\Delta f_0$ is the intrinsic contribution to the linewidth.

In the fully FM regime, $\alpha$ has been measured to be $\alpha = (1.3 \pm 0.8) \times 10^{-3}$ by Mancini \textit{et al}. \cite{mancini}. For the closest measurement to the fully FM phase here, which corresponds to $t_\text{AF} = 0.6 \pm 0.01$ nm, we see $\alpha = (23 \pm 3)\times 10^{-3}$. Again, here we see that even modest amounts of AF material have a profound effect on the high frequency behavior of the system through mechanisms such as spin pumping. This marked increase in $\alpha$ compared to the fully FM phase decreases with increasing $t_\text{AF}$, which appears to reach its minimum value when cooling between $t_\text{AF} \sim 2-4$ nm, the same value at which the peak in $A^\prime$ is seen in the SWR experiment when cooling. Interestingly, for thicknesses in excess of $t_\text{AF} \sim 4$ nm, $\alpha$ demonstrates a quasi-linear increase with $t_\text{AF}$, which is behavior attributed to the dephasing of spin-currents that are pumped into the AF material as they pass through \cite{Meridio2014}. The heating arm also demonstrates a similar behavior, though the peak in $A^\prime$ and the position of the dip in $\alpha$ do not line up exactly.

This increase in $\alpha$ with thickness compared to measured value for the fully FM phase is again indicative of a TDPT taking place within the AF layer \cite{Mccord2004,Frangou2016}. The position in which this TDPT finishes on the cooling arm is unclear, but it occurs somewhere within the $t_\text{AF} \sim 2-4$ nm, as behavior consistent with the presence of bulk AF material is seen after this thickness \cite{Meridio2014}. This is the thickness at which the PSSWs are lost in the SWR experiment when cooling. Interestingly, the peak in $A^\prime$ seen when heating also coincides with the dip seen in $\alpha$ for the same transition branch. As is the case with the cooling branch measurements, the PSSW excitations are also lost at thicknesses in excess of this dip which implies that the same behavior is seen on this transition branch also. 

However, it is unclear here why the TDPT would lead to an increase in $A^\prime$ at this stage. To gain a better understanding of the how the interphase exchange coupling affects the physical properties of the system, computer simulations of atomistic spin-dynamics were performed.

\section{Computer Simulations of Atomistic Spin-Dynamics}

Previously, the first-order phase transition in FeRh system has been modelled on the basis of the different temperature scalings of the bilinear exchange interaction and the higher-order exchange interactions between quartets of Fe atoms known as the four-spin interaction \cite{Barker2015}. This model uses the four-spin term to mediate the interactions due to the Rh moment, and it is this interaction that is responsible for the AF ordering at low temperature. The four-spin interaction is more sensitive to temperature fluctuations than the bilinear exchange and breaks down at lower temperatures. At this point, the bilinear exchange, which mediates the FM interactions between adjacent Fe atoms, takes over and the metamagnetic transition occurs. The spin Hamiltonian described in Ref.~\onlinecite{Barker2015} includes the nearest and next-nearest neighbour bilinear and four-spin interaction terms, and here includes the uniaxial anisotropy, $K_\text{u}$, such that:
\begin{eqnarray}
\label{gen_ham}
{\mathcal{H}} &=&-\frac{1}{2}\,  \sum_{i,j } J_{ij}\, (\mathbf{S}_i\, \cdot  \mathbf{S}_j\:) \nonumber -\frac{1}{3}\,  \sum_{i,j,k,l } D_{ijkl}\, (\mathbf{S}_i\, \cdot  \mathbf{S}_j\:)(\mathbf{S}_k\, \cdot  \mathbf{S}_l\:) \nonumber \\
& &\:-\sum_{i}(\mu_\text{i} \mathbf{S}_i\, \cdot  \left[\mathbf{B}_\text{Ext} + \mathbf{B}_\text{RF}\right]) -K_\text{u} \sum_{i} (\mathbf{S}_i\, \cdot  \hat{\mathbf{e}})^2,
\end{eqnarray}
where $J_{ij}$ and $D_{ijkl}$ represent the bilinear and four-spin exchange interactions between Fe atomic sites with spin $\mathbf{S}_i$ and $\hat{\mathbf{e}}$ represents the easy axis direction. We performed simulations using this Hamiltonian using the \textsc{Vampire} software package \cite{Evans2014}.

\begin{figure*}[t]
  \includegraphics[width = 17.5cm]{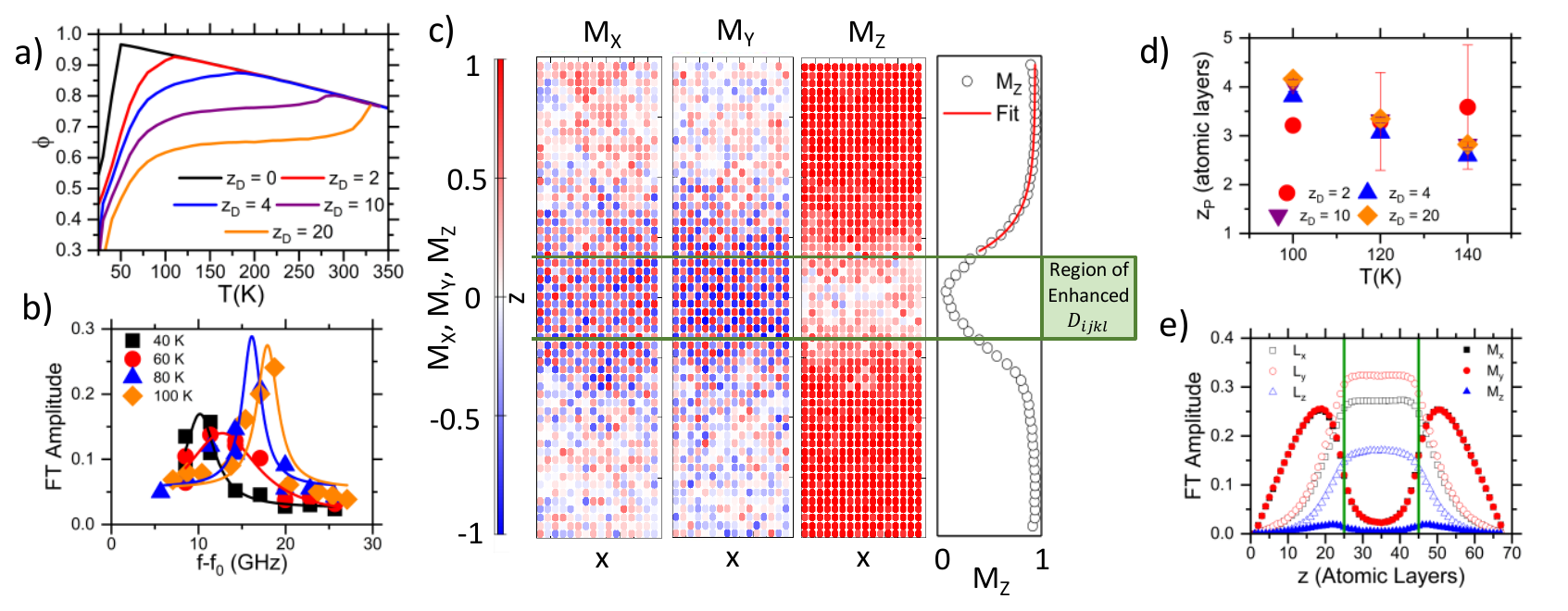}
  \caption{Atomistic simulation results. (a) Temperature dependence of the magnetization with $\mu_0 H_\text{Ext} = 2$~T for simulated systems with various $z_\text{D}$. (b) Fourier transformed first SWR mode for $z_\text{D}$ = 20 at various temperatures with a fit of a Lorentzian function to the data points (solid lines). (c) Simulation output for a $z_\text{D}$ = 10 (green region) system taken 1~ns after the field cooling has finished at a temperature of 100~K. The magnetization of the $x,y,z$ directions are shown alongside the $M_\text{Z}$ profile fitted for penetration. AF order extends beyond the boundaries of the central region by several atomic planes. (d) results of fitting for $z_\text{P}$ against temperature for various $z_\text{D}$. (e) maximum amplitude of the FT of $\mathbf{L}$ and $\mathbf{M}$ across the film for the $z_\text{D} = 20$ system at 120~K excited via an applied RF field. The region of increase $D_{ijkl}$ is also shown in green here.}
\label{fig:simsfig}
\end{figure*}

The experimental setup is replicated exactly in the simulations, with an in-plane RF field $\mathbf{B}_\text{RF} = \mu_0 \mathbf{H}_\text{RF}$ applied at frequency $\nu$ whilst a static field $\mathbf{B}_\text{Ext} = \mu_0 \mathbf{H}_\text{Ext}$ is applied perpendicular to the film plane. The spins are pinned at either end of the film thickness. The Fourier transform of the time dependence of the in-plane magnetization component is taken as the simulated SWR spectra. The thermal hysteresis associated with the first-order phase transition requires the coexistence of the two magnetic phases and ordinarily needs to be assured by a large enough system size. The system size is limited here by the computationally expensive nature of the four-spin exchange term, which means it is necessary to artificially introduce the AF regions in these systems. Our modelling is therefore not intended to fully quantitatively match the experimental results, but rather to capture the same physics in a semi-qualitative way.

To achieve this we turned again to our trilayer model, and defined a region in the middle of the system, $z_\text{D}$ atomic planes thick, that has its four-spin exchange interaction strength $D_{ijkl}$ increased to $D_{q,2}$. This creates a region with a higher $T_\text{T}$ than its surroundings, where $D_{ijkl} = D_{q,1}$. For a given temperature range this generates a FM/AF/FM trilayer geometry, just like that in Fig.~\ref{fig:trilayer}, as can be seen in Fig.~\ref{fig:simsfig}(c). $z_\text{D}$ is varied to simulate different points within the transition. This amounts to varying $t_\text{AF}$, and hence $\phi$, in the terminology of the previous section. The parameters used in the simulations are taken from Ref.~\onlinecite{Barker2015} and are presented in Table~\ref{parameters_simulation}, with the exception of the value of the $K_\text{u}$, which was taken from Ref.~\onlinecite{Ostler2017}.

\begin{table}[b!]
\caption{Simulation parameters.
\label{parameters_simulation}}
\begin{tabular}{c c}
\hline \hline
Quantity  & Value     \\ \hline
$J_{\text{nn}}$  & $0.4 \times 10^{-21}$ ~J \\
$J_{\text{nnn}}$  & $2.75 \times 10^{-21}$~J \\
$D_{q,1}$  & $0.16 \times 10^{-21}$~J \\
$D_{q,2}$  & $0.23 \times 10^{-21}$~J \\
$K_\text{u}$  & $1.404 \times 10^{-23}$~J \\
$\mu_\text{Fe}$  & $3.15$~$\mu_\text{B}$ \\
$|\mu_0 \mathbf{H}_\text{Ext}| $  & $2$~T \\
$|\mu_0 \mathbf{H}_\text{RF}|$  & $0.05$~T \\
$\nu$  & varied GHz \\
$z_\text{D}$ & 0, 2, 4, 10 \& 20 \\
$\alpha$ & 0.01 or 1 \\
\hline \hline
\end{tabular}
\end{table}

The simulated temperature dependent magnetization for each value of $z_\text{D}$ using the experimental configuration is shown in Fig.~\ref{fig:simsfig}(a). Adding the intermediate region of enhanced four-spin interactions gives a wide range of temperatures where the MMP regime exists, yielding broad transitions qualitatively comparable to the experimental sample.

The Fourier-transformed first SW modes for the $z_\text{D} = 20$ system, shown in Fig.~\ref{fig:simsfig}(b), show a reduction in resonant frequency and amplitude with falling temperature, consistent with the experimental results. The quantitative discrepancies between experiment and the simulations for the SW modes is due to differences in thickness, with the difference in temperatures due to the smaller system sizes used in the simulations. In this experiment the value of $\alpha = 0.01$ is used. To look for evidence of the interphase exchange coupling, field cooled simulations were performed and the magnetic structure through the stack was visualised.

The simulated system was field cooled under $\mu_0 H_{\text{Ext}} = 2$~T from 750~K to either 100, 120, or 140~K for 1~ns and then allowed to evolve for the same time again, by which time the system had equilibrated. To help aid the quick settling of the state, the value of the Gilbert damping parameter was set to the artificially high value of $\alpha = 1$. An example of this final state for the $z_\text{D} = 10~$ system is shown in Fig.~\ref{fig:simsfig}(c). In this figure, the colour in the stack represents the direction and strength of the magnetization at each point. The colour scale used is shown on the left-hand side of panel (c). The FM order along the $+z$ direction of $\mu_0 H_{\text{Ext}}$ is evident either side of the region encompassed by $z_\text{D}$. Within that region, shown in green here, in-plane AF order is present. As the AF order is known to orient itself perpendicularly to the applied field to minimize the energy in this system \cite{Marti2014}, this configuration is expected. It also demonstrates the recovery of the trilayer configuration using varying values of $D_{ijkl}$ for the different magnetic phases. Just outside the region of enhanced $D_{ijkl}$, there is an ambiguous spin structure that exhibits aspects of both AF and FM order.

One would na\"{\i}vely expect that of the sixty layers in total, only the ten layers with an enhanced four-spin interaction would be in the AF phase, leading to a $\phi \sim 0.83$. From the magnetization profiles in Fig.~\ref{fig:trilayer}(a), it can be seen that the system actually exhibits $\phi \sim 0.7$ at this temperature, equivalent to 18 layers being in the AF phase. We ought therefore to expect to find the equivalent of four layers of full AF order on each side of the region described by $z_\text{D}$. Consulting Fig.~\ref{fig:simsfig}(c) we can see that in fact this region is smeared out over a greater distance with partial FM and AF order. The fully AF region contained within $z_\text{D}$ has a profound effect on the notionally FM material around it, overcoming to a considerable extent the bilinear exchange interaction. This is consistent with our experimental observation that even a small departure of $\phi$ from 1 leads to a marked effect on the measured exchange stiffness.

In order to quantitatively analyse the extent of this region, the average magnetization per layer starting from the region encompassed by $z_\text{D}$ is fitted to $f(z) = M_\text{S} - \Delta \exp ({-\frac{z}{z_\text{P}}})$, where $z$ is the layer index, $\Delta$ is the amplitude of magnetization weakening and $z_\text{P}$ represents the lengthscale on which FM order is recovered. An example of this fitting is seen in Fig.~\ref{fig:simsfig}(c) with the average value of $z_\text{P}$ extracted for a selection of time-steps taken after the field-cooling protocol has finished is shown in Fig.~\ref{fig:simsfig}(d). These averages are calculated assuming that everything outside of a 95\% confidence level is an outlier, which are not considered in the calculation. The fit shown in Fig.~\ref{fig:simsfig}(c) is not used in the calculation of the average and is shown as a representative example.  

The results for $z_\text{P}$ show that between 2 and 5 atomic layers in each of the regions outside $z_\text{D}$ have their magnetic order changed from fully FM by proximity to the enhanced four-spin interaction region, depending on temperature and $z_\text{D}$. The saturation value of $z_\text{P}$ corresponds to a thickness of $1.25 \pm 0.02$ nm, which agrees reasonably well the position of the plateau seen in the behavior of $A^\prime$ for the cooling branch measurements. This then adds credence to the idea that the behavior of $A^\prime$ measured in the experiment can be interpreted as a development of an exchange coupling across the AF/FM interface with increasing $t_\text{AF}$. The development of this exchange coupling with thickness is attributed to a TDPT that takes place in the AF layer: below a critical thickness set by $z_\text{P}$ there will be no fully AF region within the middle of the nucleus that is making a transition out of the fully AF state. the appearance of this full AF order is a candidate for the TDPT that is analogous to the one that occurs in exchange bias systems.

From the behavior of $A^\prime$ and $\alpha$ seen in the experiment, the TDPT finishes between $t_\text{AF} \sim 2 - 4$ nm, which is in excess of the plateau associated with the establishment of the exchange coupling. From the simulations of the SWR modes in Fig.~\ref{fig:simsfig}(b) it is known that the PSSWs travel through the AF layer in order to give modes that span the total thickness of the layer. The experiment also suggests that the PSSWs pass through bulk AF material so long as it is sufficiently thin: but cannot penetrate more than several nm since that is when the SWR resonances are lost in the experiment. Therefore, in order to identify how these excitations travel through the AF layer in the experiment, investigations into the development of the spin-wave transmission through the AF layer with thickness were made.

In AF/FM bilayers that are driven at the resonant frequency of the FM layer, theory predicts that the exchange coupling at the interface can excite collective excitations of the Ne\'el vector, $\mathbf{L} = (\mathbf{m}_1 - \mathbf{m}_2)/2 M_\text{S}$, with $\mathbf{m}_i$ denoting the AF sublattice magnetization in the AF layer \cite{Khymyn2016}. Due to the difference in the resonant frequencies of the two layers, these excitations of the Ne\'el vector travel only a finite distance through the AF layer and are therefore known as evanescent SWs (ESWs) \cite{Khymyn2016}. These ESWs have been seen to carry spin-currents through distances up to 10~nm in the AF insulator NiO \cite{Wang2014,Khymyn2016} and so could be responsible for the transmission of the PSSW through the AF layer seen in this experiment down to the temperature at which $t_\text{AF}$ become too large and the connection between the FM layers, and hence the PSSW mode, is lost. Fig.~\ref{fig:simsfig}(e) shows the amplitude of the Fourier transform for each of the directional components of both $\mathbf{L}$ and the magnetization vector $\mathbf{M} = (\mathbf{m}_1 + \mathbf{m}_2)/2 M_\text{S}$ through the stack, in the simulated $z_\text{D}$ = 20 system at 120 K. The value of $\alpha = 0.01$ is again used in this experiment. Here, it is clear that the amplitude of the Fourier transform of $\mathbf{L}$ is non-zero for all directional components in the region defined by $z_\text{D}$, as are the corresponding values of $\mathbf{M}$. This implies that there is a coherent rotation of the Ne\'el order within the bulk AF material that carries the PSSW excitation through the AF layer. As the simulations treat all materials as electrical insulators and do not consider the contribution of electrons \cite{Evans2014}, the observation of ESWs and excitations of PSSWs in the simulations confirm that this is a mechanism by which the PSSW can pass through the AF layer. Also, as the ESWs require an exchange coupling between the AF/FM boundary, it also constitutes proof of the existence of a coupling between the two layers in the simulation, consistent with that seen in the experiment in Fig.~\ref{fig:trilayer}(b).

In the experimental situation however, we do not have exactly this situation since both magnetic phases of FeRh are metallic \cite{devries2013}, and spin-currents could pass through the AF layer via electrons as well as the magnon flows that occur in insulators. In the thickness regime in which the TDPT occurs, there will be little magnon spin-pumping due to the lack of a well-defined Ne\'el vector in the AF layer. However, after a given thickness when regions of stable AF material with a common Ne\'el vector are present, ESWs could be excited in this system. As the increase in $A^\prime$ occurs in the thickness regime in which the TDPT is close to its conclusion from the behavior of $\alpha$, this peak in $A^\prime$ may then correspond to the point at which ESWs are excited in the experiment. The introduction of ESWs into the system will provide another channel through which the PSSW can travel through the AF layer and will increase the exchange energy within this layer and therefore $A^\prime$. As these ESWs by nature only travel a finite distance through the AF layer, it is believed that the disappearance of $A^\prime$ on both transition branches corresponds to the thickness in which the ESWs can no longer pass through the AF layer.

The PSSW excitations are seen for much larger $t_\text{AF}$ on the heating branch of the transition when compared to the cooling branch. The critical thickness, $t_\text{C}$, at which these TDPT take place is determined by the strength of the exchange coupling and the anisotropy energy of the AF layer, $K_\text{AF}$, such that $t_\text{C} = A_\text{Int}/2K_\text{AF}$ \cite{Radu2005}. The discrepancy in $t_\text{AF}$ where the TDPT takes place between the transition branches then implies either: the exchange coupling is higher on the heating branch than it is when cooling, or that the anisotropy energy of the AF layer is much lower when heating compared to cooling. It has been previously demonstrated that the anisotropy energy of both the FM and AF phases are largely invariant through the range of the transition probed here \cite{Clarkson2017}, which implies that the difference in the thickness range of the TDPT is due to the size of the interlayer exchange coupling. A higher value of the interlayer exchange coupling when heating compared to cooling would explain the asymmetry seen between the transition branches for the extracted values of $A$ seen in Fig.~\ref{fig:A}(b) and the increase in the baseline upon which the peak in $A^\prime$ is situated in Fig.~\ref{fig:trilayer}(a), as well as the larger the observation of PSSW excitations through larger $t_\text{AF}$ seen in the experimental data in Fig.~\ref{fig:trilayer}. The reason for this larger exchange coupling when heating is unclear at this stage. It is also not possible to explain the peak in $A^\prime$ seen in this transition branch, as it falls outside the thickness regime where the TDPT is expected and does not align with the dip in $\alpha$ extracted from the FMR measurements.

\section{Conclusion}

To conclude, SWR measurements of a Pd-doped FeRh epilayer taken during the metamagnetic transition reveal a reduced value of the measured exchange stiffness when compared to the value for the fully FM phase. By introducing the trilayer model it is possible to demonstrate that this suppression of the measured value of the exchange stiffness is due to the contribution of the AF material. The AF layer contributes an effective exchange stiffness, $A^\prime$, to the measured value. $A^\prime$ is seen to be highly dependent on the thickness of the AF layer, increasing to a plateau initially and then peaking for measurements performed when cooling. The development of the $A^\prime$ with $t_\text{AF}$ is then consistent with the development of an exchange coupling across the phase boundary due to a TDPT in the AF layer. Evidence for the presence of a TDPT in the AF layer is also provided by the behavior of the Gilbert damping parameter, $\alpha$, calculated from the FMR measurements. A peak is also seen on top of a flat signal for the heating branch.

To identify the origin of these features complementary simulations of atomistic spin-dynamics were performed using the same geometry and conditions as the experiment. The AF phase fraction was controlled by introducing a thin layer of variable thickness $z_\text{D}$ in which the strength of the four-spin interaction was enhanced relative to its surroundings. A similar reduction of SWR mode frequency with temperature was observed in the simulations. These simulations reveal that the exchange coupling across the magnetic phase boundary introduces regions of weakened FM exchange in the AF region, implying that the TDPT observed in the experiment takes place between a disordered intermixed phase in the AF layer, to a globally order bulk-like AF phase. The simulations also reveal that ESWs are present that carry the excitation through bulk AF material. Again, by comparing the behavior of $A^\prime$ and $\alpha$, it is possible to attribute the peak seen in $A^\prime$ to the introduction of ESWs into the system as the TDPT in the AF layer finishes. The properties of the exchange coupling are also seen to be asymmetric depending on the temperature sweep direction.

The influence of the interphase exchange coupling and its development through the transition with the thickness of the AF layer should be considered in the design of FeRh-based spintronic devices, particularly those intended to operate at GHz frequencies.

\begin{acknowledgments}
J.M., K.M. and M.S. contributed equally to this work. This work was supported by the Diamond Light Source, the UK EPSRC (grant number EP/M018504/1), the Advanced Storage Research Consortium and a Grant-in-aid for Scientific Research on Innovative Area, ``Nano Spin Conversion Science'' (Grant No. 26103002).
\end{acknowledgments}	

%

\end{document}



\title{Phase Boundary Exchange Coupling in the Mixed Magnetic Phase Regime of a Pd-doped FeRh Epilayer: \\ Supplementary Information} 

\author{J.~R.~Massey}
\email{J.R.Massey@leeds.ac.uk}
\affiliation{School of Physics and Astronomy, University of Leeds, Leeds LS2 9JT, United Kingdom.}

\author{K.~Matsumoto}
\affiliation{Institute for Solid State Physics, University of Tokyo, Kashiwa, Chiba 277-8581, Japan.}

\author{M.~Strungaru}
\affiliation{Department of Physics, University of York, York YO10 5DD, United Kingdom.}

\author{R.~C.~Temple}
\affiliation{School of Physics and Astronomy, University of Leeds, Leeds LS2 9JT, United Kingdom.}

\author{T.~Higo}
\affiliation{Institute for Solid State Physics, University of Tokyo, Kashiwa, Chiba 277-8581, Japan.}

\author{K.~Kondou}
\affiliation{Center for Emergent Matter Science, RIKEN, Wako, Saitama 351-0198, Japan.}

\author{R.~F.~L.~Evans}
\affiliation{Department of Physics, University of York, York YO10 5DD, United Kingdom.}

\author{G.~Burnell}
\affiliation{School of Physics and Astronomy, University of Leeds, Leeds LS2 9JT, United Kingdom.}

\author{R.~W.~Chantrell}
\affiliation{Department of Physics, University of York, York YO10 5DD, United Kingdom.}

\author{Y.~Otani}
\affiliation{Institute for Solid State Physics, University of Tokyo, Kashiwa, Chiba 277-8581, Japan.}
\affiliation{Center for Emergent Matter Science, RIKEN, Wako, Saitama 351-0198, Japan.}

\author{C.~H.~Marrows}
\email{c.h.marrows@leeds.ac.uk}
\affiliation{School of Physics and Astronomy, University of Leeds, Leeds LS2 9JT, United Kingdom.}

\date{\today}

\maketitle
\section{Temperature Dependence of the Exchange Stiffness At Various Magnetic Field Strengths}

\begin{figure}[t!]
  \includegraphics[width = 16cm]{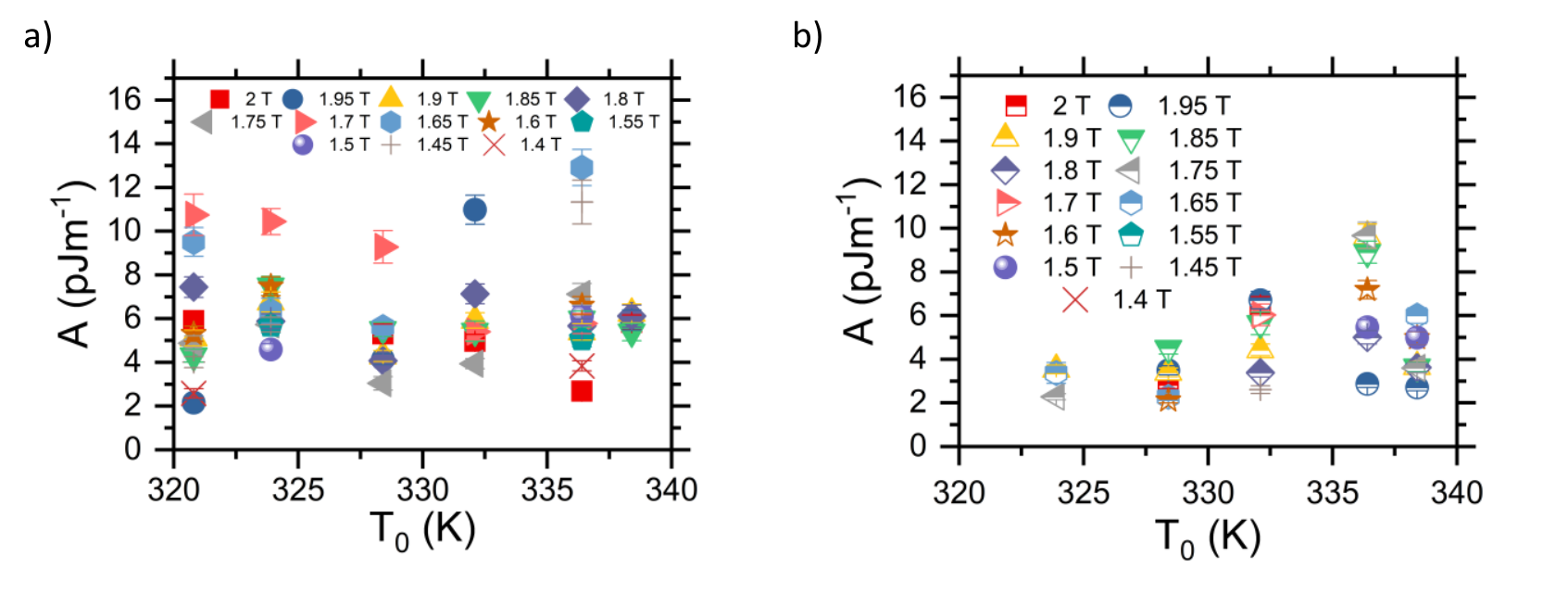}
  \caption{Temperature dependence of $A$ at different magnetic fields. (a) The extracted values of $A$ for all measurements taken when cooling plotted against the measured temperature at which they are extracted, $T_0$, for all values of the applied field, $\mu_0 H_\text{Ext}$, used in the experiment. (b) The same quantities as plotted in panel (a) but for measurements performed when heating. The different colours are used to show measurements performed at different magnetic fields with the index listed in the key in each figure. This figure shows that there is no consistent trend through the temperature range for a given magnetic field when cooling, whereas the heating branch measurements show a peak for all fields at $T_0 \sim 337$ K. \label{fig:Supp_A}}
\end{figure}

Fig.~\ref{fig:Supp_A} shows the dependence of the extracted exchange stiffness, $A$, on the measured temperature, $T_0$, at various magnetic field strengths through the temperature range available in this experiment, for measurements performed when cooling (panel (a)) and heating (panel (b)). The measurements performed when cooling do not demonstrate a clear trend through the temperature range for any externally applied field. This is consistent with the raw data shown against the effective temperature in the Fig.~3(a) of the main text. The measurements performed when heating however show a peak in $A$ at $T_0 \sim 337$ K. This corresponds to the effective temperature range of $T_\text{Eff} \sim 350 - 355$ K where the peak is seen in Fig.~3(a) of the main text.

%